\documentclass[12pt]{article}
\usepackage{epsfig}

\def\beq{\begin{equation}}
\def\eeq#1{\label{#1}\end{equation}}
\def\eeqn{\end{equation}}

\def\beqa{\begin{eqnarray}}
\def\eeqa#1{\label{#1}\end{eqnarray}}
\def\eeqan{\end{eqnarray}}

\let\bar=\overbar

\def\Dslash{\not{\hbox{\kern-4pt $D$}}}
\def\dslash{\not{\hbox{\kern-2pt $\del$}}}

\def\msb{{\bar{\ssstyle M \kern -1pt S}}}

\usepackage{fancyhdr,graphicx}
\def\Title#1{\begin{center} {\Large {\bf #1} } \end{center}}

\begin{document}

\Title{Neutrino and Electron-positron Pair Emission from Phase-induced Collapse of Neutron Stars to Quark Stars}

\bigskip\bigskip


\begin{raggedright}

{\it K. S. Cheng\index{Cheng, K. S.}\\
Department of Physics and Center for Theoretical and Computational Physics, \\
The University of Hong Kong, Pok Fu Lam Road, Hong Kong, Hong Kong SAR, \\
P. R. China \\
{\tt Email: hrspksc@hkucc.hku.hk}}
\bigskip\bigskip
\end{raggedright}

\begin{raggedright}

{\it T. Harko\index{Harko, T.}\\
Department of Physics and Center for Theoretical and Computational Physics, \\
The University of Hong Kong, Pok Fu Lam Road, Hong Kong, Hong Kong SAR, \\
P. R. China \\
{\tt Email: harko@hkucc.hku.hk}}
\bigskip\bigskip
\end{raggedright}

\begin{abstract}
We study the energy released from
phase-transition induced collapse of neutron stars, which results in
large amplitude stellar oscillations. To model this process we use a Newtonian hydrodynamic code, with a high resolution
shock-capturing scheme. The physical
process considered is a sudden phase transition from
normal nuclear matter to a mixed phase of quark and nuclear matter.
We show that both the temperature and the density at the
neutrinosphere oscillate with time. However, they are nearly
180$^{\circ}$ out of phase. Consequently, extremely intense,
pulsating neutrino/antineutrino and leptonic pair fluxes will be
emitted. During this stage several mass ejecta can be ejected from
the stellar surface by the neutrinos and antineutrinos. These ejecta
can be further accelerated to relativistic speeds by the
electron/positron pairs, created by the neutrino and antineutrino
annihilation outside the stellar surface. We suggest that this process
may be a possible mechanism for short Gamma-Ray Bursts.
\end{abstract}

\section{Introduction}

Recently, by using simulations performed with the
Newtonian numerical code introduced in \cite{Lin06}, it was shown that the resulting quark star, produced by
the phase transition induced collapse of a neutron star, will undergo a series of oscillations.  The collapse process with a
conformally flat approximation to general relativity was also simulated in \cite{Ab08}. The works of
\cite{Lin06,Ab08} focus on the gravitational wave signals emitted
by the collapse process.

It is the purpose of the present paper to
consider another important implication of this result, namely, the
effect of the oscillations of the newly formed quark star on the
neutrino emission. The oscillations can enhance the neutrino
emission rate in a pulsating manner, and the neutrinos are emitted
in a much shorter time scale. Moreover, through the process of neutrino-antineutrino annihilation, a large amount of electron-positron pairs is also produced. Therefore the neutron-quark phase
transition in compact objects may be the energy source of GRBs \cite{Ch09}. Such a model can also explain the lack of detection of a neutron star or pulsar formed in the SN 1987A \cite{Cha09}, by assuming that the newly formed neutron star at the center of SN 1987A underwent a phase transition after the neutrino trapping timescale ($\sim$ 10 s). Consequently,  the compact remnant of SN 1987A may be a strange quark star, which has a softer equation of state than that of neutron star matter. Such a phase transition can induce stellar collapse and result in large amplitude stellar oscillations. Extremely intense pulsating neutrino fluxes, with submillisecond period and with neutrino energy (greater than 30 MeV), can be emitted because the oscillations of the temperature and density are out of phase almost 180 $^{\circ}$. If this is true, the current X-ray emission from the compact remnant of SN 1987A will be lower than $10^{34}$ erg s$^{-1}$, and it should be a thermal bremsstrahlung spectrum for a bare strange star with a surface temperature of around $\sim $ 107 K.

\section{Description of the phase transition and of the equations of state}

The numerical code is based on the three-dimensional numerical
simulation in Newtonian hydrodynamics and gravity. The quark matter
of the mixed phase is described by the MIT bag model and the normal
nuclear matter is described by an ideal fluid EOS. This code has
been used to study the gravitational wave emission from the
phase-induced collapse of the neutron stars \cite{Lin06}, where a detailed discussion can be found. In the simulations, we introduced a very low density atmosphere
outside the star. The ''artificial'' atmosphere is not physical, but
it is important  for the stability of the  hydrodynamical code. This
is due to the problem that the hydrodynamical codes cannot in
general  handle vacuum  regions where the density is zero. In order
to avoid a significant influence of the atmosphere on the dynamics
of the physical system, it is necessary to choose the density of the
atmosphere $\rho_{\rm atm}$ to be much smaller than the density
scale of interest. For the results reported in this paper, we set
$\rho_{\rm atm}$ to be $3\times 10^9 \;{\rm g/cm}^3$.

In our simulations, we will not simulate the phase transition process. Instead,
we assume that a fast phase transition has happened (e.g., via a detonation
mode) so that the initial neutron star has converted to a quark star in a
timescale shorter than the dynamical timescale of the system.
We assume that the normal matter inside the initial
neutron star has suddenly changed to quark matter at $t=0$. This is achieved
by changing the EOS at $t=0$ after the initial hydrostatic equilibrium neutron
star has been constructed. We then simulate the resulting dynamics of the system
triggered by the collapse.

The equation of state (EOS) for neutron stars is highly uncertain.
We could try all possible existing realistic EOS in our study.
However, the main purpose of this paper is to demonstrate that
during the phase-transition induced collapse of a neutron star
extremely intense, pulsating and very high energy neutrinos can be
emitted. The effect is governed mainly by the amount of pressure
reduction after the phase transition as compared to the initial
neutron star model. For simplicity we will use a polytropic EOS for
the initial equilibrium neutron star:
\begin{equation}
P=k_0\rho^{\Gamma_0} , \label{eq:poly_EOS}
\end{equation}
where $k_0$ and $\Gamma_0$ are constants. On the initial time slice,
we also need to specify the specific internal energy $\epsilon$. For
the polytropic EOS, the thermodynamically consistent $\epsilon$ is
given by
\begin{equation}
\epsilon = {k_0\over \Gamma_0-1} \rho^{\Gamma_0-1} .
\end{equation}
Note that the pressure in Eq.~(\ref{eq:poly_EOS}) can also be
written as
\begin{equation}
P=(\Gamma_0-1)\rho\epsilon .
\end{equation}

In this paper we use a mixed phase EOS to mimic a quark star core
covered by normal nuclear matter. This EOS model consists of two
parts: (i) a mixed phase of quark and nuclear matter in the core at
density higher than a certain critical value $\rho_{tr}$ (quark
seeds can spontaneously produce everywhere when $\rho \geq \rho
_{tr}$) (ii) a normal nuclear matter region extending from $\rho <
\rho_{tr}$ to the surface of the star. A more detailed discussion
about such hybrid quark stars can be found in \cite{Lin06}.
Explicitly, the pressure is given by
\begin{equation}
P = \left\{ \begin{array}{cc}
           \alpha P_{\rm q} + (1-\alpha) P_{\rm n},
         & \ \mbox{for} \ \  \rho > \rho_{tr} \\
        \\
            P_{\rm n},
         & \mbox{for} \ \ \rho  \leq \rho_{tr} ,  \end{array}  \right.
\label{eq:mixed_EOS}
\end{equation}
where
\begin{equation}
P_{\rm q} = {1\over 3} \left( \rho + \rho\epsilon - 4B\right),
\label{eq:quark_eos}
\end{equation}
is the pressure contribution of the quark matter,
\begin{equation}
P_{\rm n}=(\Gamma_n-1)\rho \epsilon , \label{eq:eos_idealgas}
\end{equation}
 and
\begin{equation}
\alpha = \left\{ \begin{array}{cc}
          { (\rho -\rho_{tr}) / (\rho_{q} - \rho_{tr}) },
         & \ \mbox{for} \ \  \rho_{tr} < \rho < \rho_{\rm q}, \\
       \\
            1,
         & \mbox{for} \ \ \rho_{q} < \rho ,  \end{array}  \right.
\label{eq:alpha}
\end{equation}
is defined to be the scale factor of the mixed phase \cite{Lin06}.
$\Gamma_n$ is not necessarily equal to $\Gamma_0$. As for the quark
matter we have assumed that it is described by the MIT
bag model. It should be noticed that $P_q$ is not in the usual form
of $P = (\rho_{\rm tot} - 4 B)/3$, where $\rho_{\rm tot}$ is
the (rest frame) total energy density, and $B$ is the bag constant.
It is because in our Newtonian simulations, we use the rest mass
density $\rho$ and specific internal energy $\epsilon$ as
fundamental variables in the hydrodynamics equations.

The total
energy density $\rho_{\rm tot}$, which includes the rest mass
contribution, is decomposed as $\rho_{\rm tot} = \rho +
\rho\epsilon$. We choose $\Gamma_n < \Gamma_0$ in our simulations to
take into account the possibility that the nuclear matter may not be
stable during the phase transition process, and hence some quark
seeds could appear inside the nuclear matter, or the convection,
which can occur during the phase transition process, can mix some
quark matter with the nuclear matter. In the presence of the quark
seeds in the nuclear matter, the effective adiabatic index will be
reduced. The possible values of $B^{1/4}$ range from 145 MeV to 190
MeV \cite{DeGr75,Sa82,Ste95,Ha03}. For
$\rho
>\rho_q$, the quarks will be deconfined from nucleons. The value of
$\rho_q$ is model dependent; it could range from 4 to 8 $\rho_{nuc}$
\cite{Ha03, ChDa98a, Bo04}, where
$\rho_{nuc}=2.8\times 10^{14}~{\rm g~cm}^{-3}$ is the nuclear
density.

\begin{figure}[htb]
\begin{center}
\epsfig{file=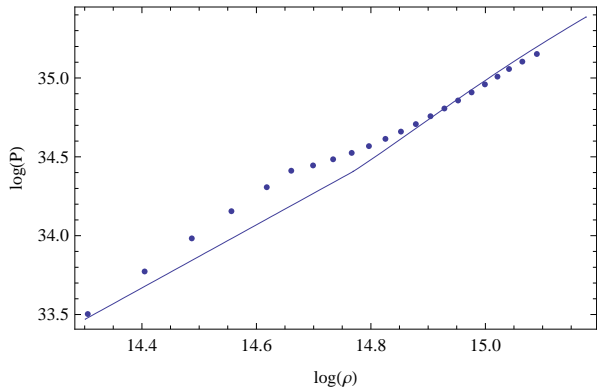,height=2in}
\caption{Comparison of the equation of state of the mixed quark-hadron phase proposed in \cite{Gl00} for $K=240$ MeV and effective mass $m^*/m=0.78$ (dotted curve), and Eq.~(\ref{eq:mixed_EOS}), the equation of state if the mixed phase used in the present simulations (solid curve).}
\label{pres}
\end{center}
\end{figure}

The properties of a mixed quark-hadron phase and its implications for hybrid star structure were considered in \cite{Gl00}. In this study it was assumed that charge neutrality must be enforced only as an overall constraint, an not separately on each of the phases. In the mixed phase the fraction $\chi $ of the volume occupied by the quark phase can be taken as the basic independent variable, being more convenient than the baryon density. The pressures of the two phases in equilibrium are equal, but vary with the quark fraction. Once the properties of the two phases are known, with the use of the relationships between the quark chemical potentials and the independent conserved quantities like baryon number and electron charge, the equation of state of the matter in the mixed phase can be obtained. In Fig.~\ref{pres} we present the equation of state of the mixed phase obtained in \cite{Gl00} for a compression modulus $K=240$ MeV and an effective mass at saturation density of $m^*/m=0.78$, and the equation of state Eq.~(\ref{eq:mixed_EOS}) used in the present simulation, respectively.

In the simulations, we choose $\Gamma_0 = 2$, $\Gamma_n = 1.85$,
$B^{1/4} = 160$ MeV and $\rho_q = 9 \rho_{\rm nuc}$. The transition
density $\rho_{tr}$ is defined to be at the point where $P_q$
vanishes initially.


\section{Emission of neutrinos and $e^{\pm}$ pairs}

The neutrino luminosity is given by \cite{Balantekin2005},
\begin{equation}
L_\nu = 4 \pi r^2 c \, \frac{1}{2 \pi \hbar ^2} \int \, \frac{E_\nu \, d^3 \mathbf{p}_\nu}{1+\exp ( E_\nu-\mu_\nu/kT_\nu)}, 
\end{equation}
where $\mu _\nu$ is the neutrino chemical potential. Taking $\mu
_\nu=0$, the neutrino luminosity emitted from the neutrinosphere is
given by $L_\nu= 7\pi R_\nu^2 ac T_\nu^4/16$, where $a=4\sigma /c$
is the radiation constant, $\sigma $ is the Stefan-Boltzmann
constant, and $T_\nu$ is the temperature of the neutrinosphere. If
we assume equal luminosities for neutrinos and antineutrinos, the
combined luminosity for a single neutrino flavor is
\begin{equation}\label{eq:lum}
L_{\nu, \, \overline{\nu}} = L_\nu + L_{\overline{\nu}}= \frac{7}{8}
\pi R_\nu^2 ac T_\nu^4.
\end{equation}

The effect of coherent forward scattering must be taken into account when considering the oscillations of neutrinos traveling through matter \cite{Wo78}. Although different flavor neutrinos have different $R_\nu$, yet they have approximately the same value
of luminosity for all flavors \cite{Janka1995,Janka2001}. Therefore the total luminosity is
around three times of a single neutrino flavor luminosity
\begin{equation}\label{eq:lumAll}
L = L_{\nu_e,\,\overline{\nu}_e} +
L_{\nu_\mu,\,\overline{\nu}_\mu} +
L_{\nu_\tau,\,\overline{\nu}_\tau}
= \frac{21}{8} \pi R_\nu^{2} ac T_\nu^4.
\end{equation}

\begin{figure}[htb]
\begin{center}
\epsfig{file=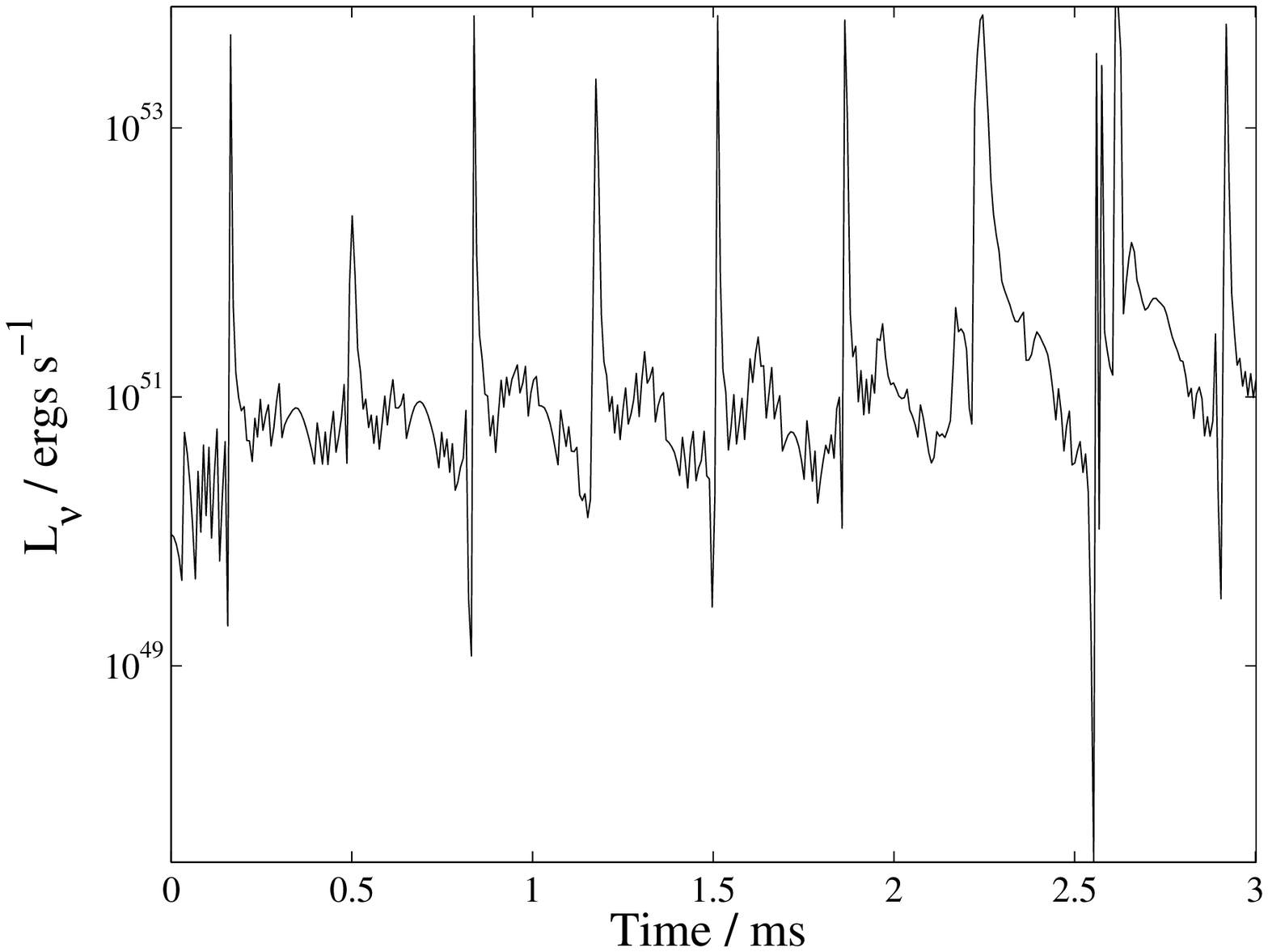,height=2in}
\epsfig{file=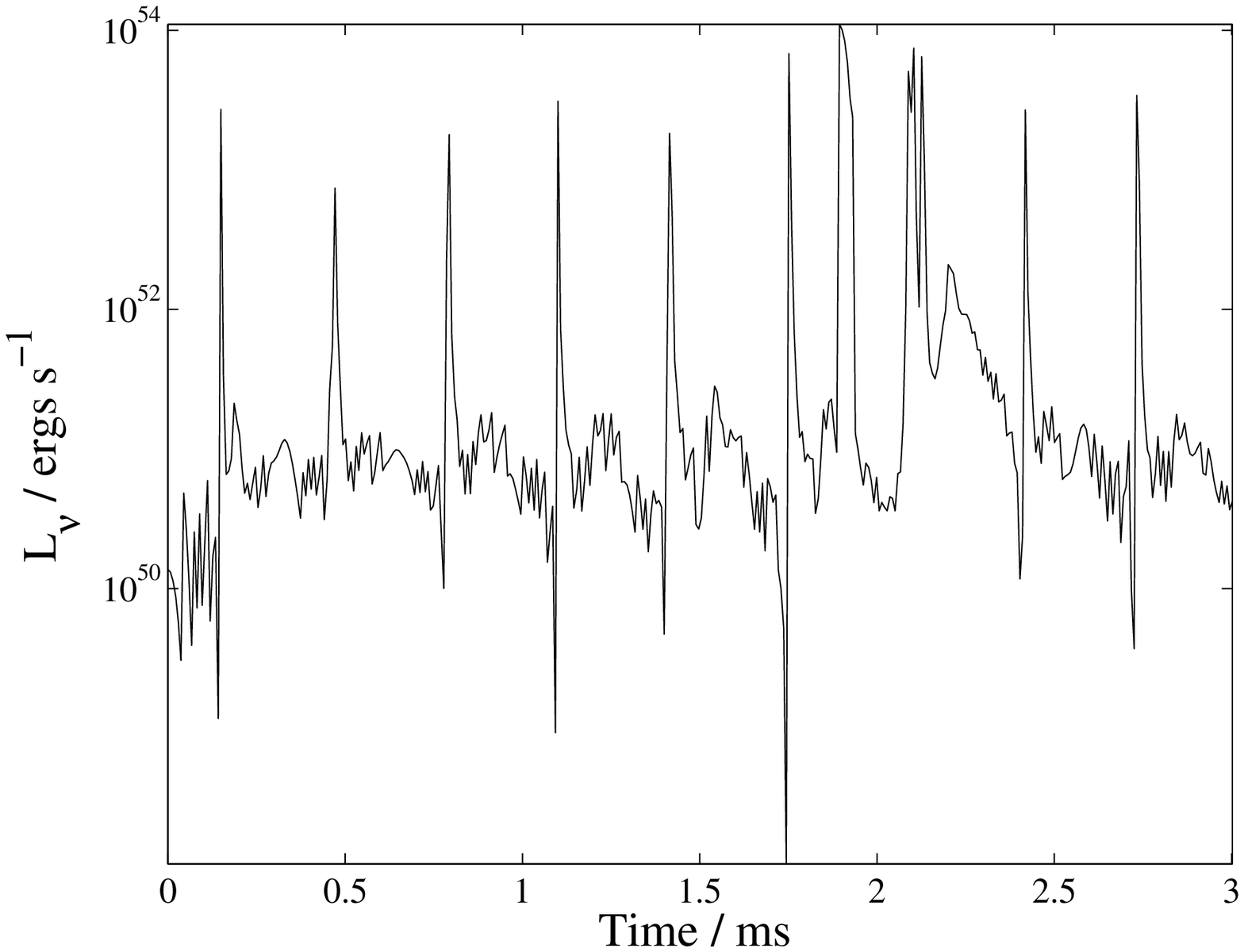,height=2in}
\epsfig{file=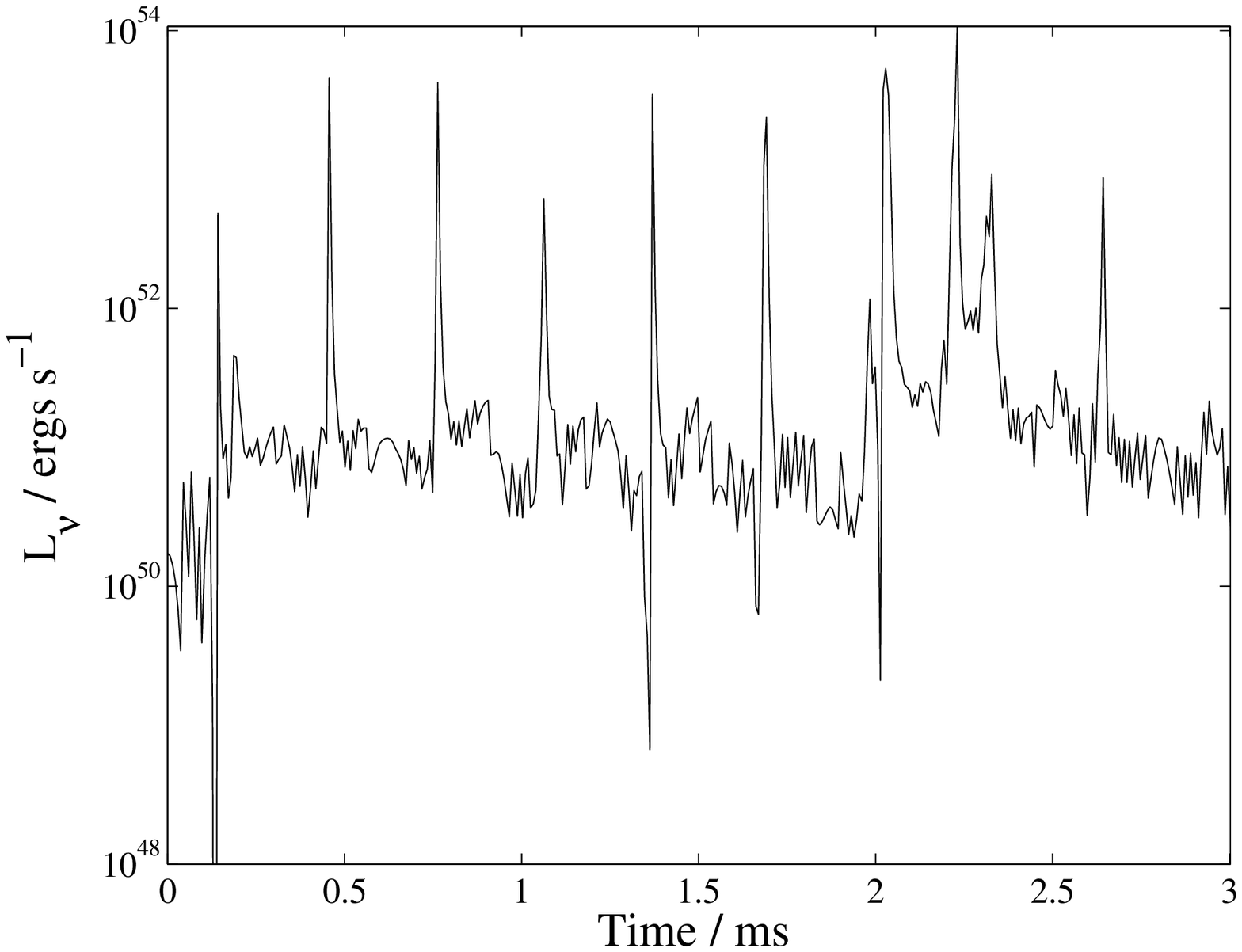,height=2in}
\epsfig{file=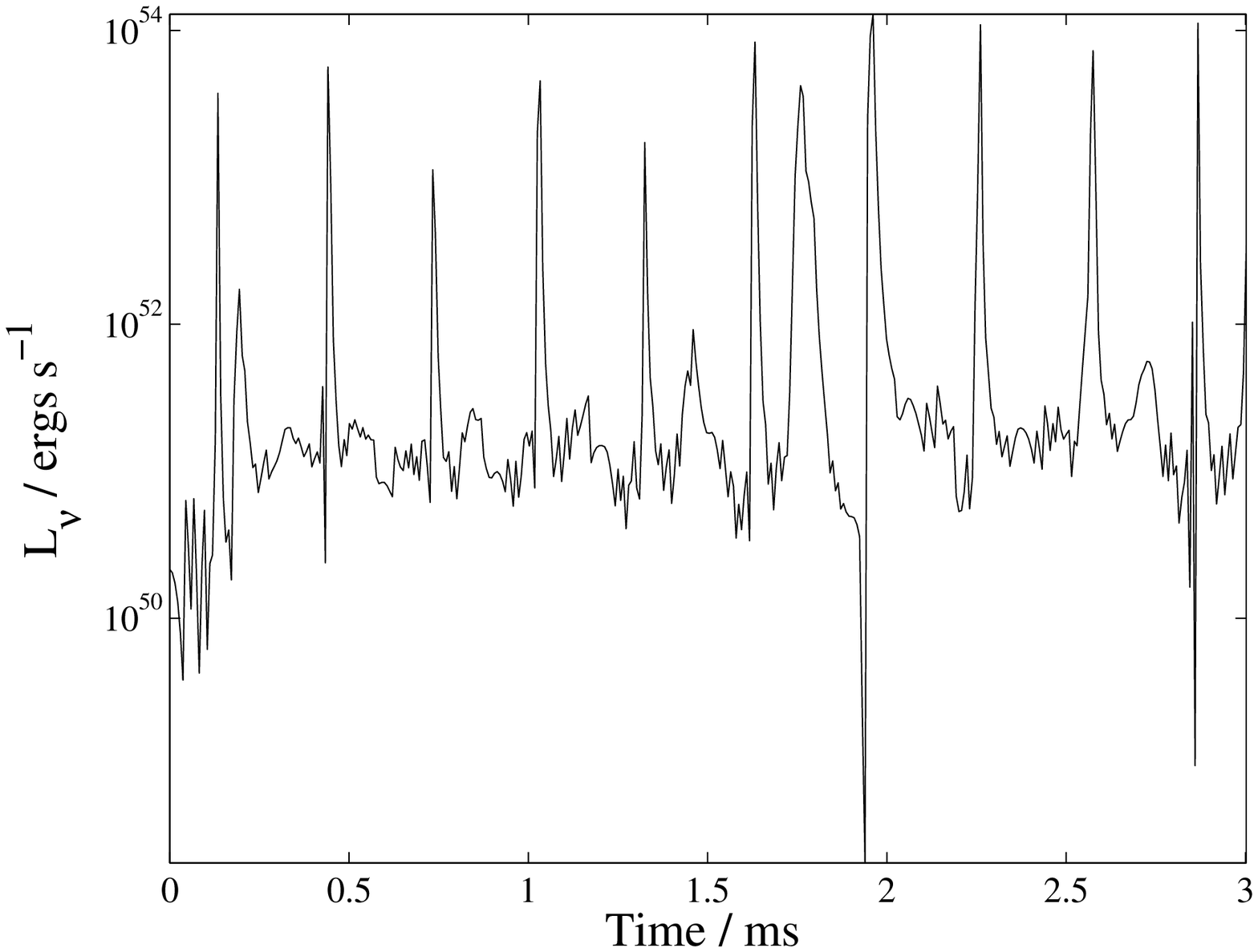,height=2in}
\caption{Neutrino luminosity versus time for $M=1.55M_{\odot}$, (left upper figure), $M=1.7M_{\odot}$, (right upper figure), $M=1.8M_{\odot}$, (left lower figure), $M=1.9M_{\odot}$, (right lower figure).}
\label{fig:Lnu_time_models}
\end{center}
\end{figure}

Using $R_\nu$ and $T_\nu$ obtained from the last  Section, we
compute the the neutrino luminosity as a function of time. The
results are shown in Fig.~\ref{fig:Lnu_time_models}.

The peak luminosities range
from $10^{52}$ to $10^{54}$ ergs/s; the pulsating period of the
luminosity is the same as that of the temperature and of the
density.

Neutrinos and antineutrinos can become electron and positron pairs
via the neutrino and antineutrino annihilation process
($\nu~+~\bar{\nu}~\rightarrow ~e^-~+~e^+ $). The total neutrino and
antineutrino annihilation rate can be given as follows \cite{Goodman1987,Cooperstein1987}
\begin{eqnarray}
\dot{Q}_{\nu \bar{\nu} \rightarrow e^{\pm}}&=&\frac{7 G_F^2 \pi^3
\zeta(5)}{2 c^5 h^6} D \left[kT_\nu(t)\right]^9 \int _{R_{\nu}} ^{\infty}
\Theta(r) \, 4 \pi r^2 dr\\ &=& \frac{7 G_{F}^{2} D \pi^3
\zeta(5)}{2 c^5 h^6} \frac{8\pi^3 }{9} R_{\nu}^3(kT_\nu)^9,
\end{eqnarray}
where $\Theta(r)=2\pi^2(1-x)^4(x^2+4x+5)/3$, $x =
\sqrt{1-R_{\nu}^2/r^2}$, $T_\nu(t)$ is the temperature at the
neutrinosphere at time $t$, $G_F^2=5.29\times 10^{-44}$ is the Fermi
constant,  $\zeta $ is the Riemann zeta function, and $D$ is a
numerical value depending on the pair creation processes (e.g.
experimental results indicate that $D_1=1.23$ for
$\nu_e\,\nu_{\bar{e}}$ and $D_2=0.814$ for $\nu_\mu \,
\nu_{\bar{\mu}}$ and $\nu_\tau \, \nu_{\bar{\tau}}$). To obtain the
total neutrino annihilation rate from all species,
$\nu_{e}\,\nu_{\bar{e}}$, $\nu_{\mu}\,\nu_{\bar{\mu}}$ and
$\nu_{\tau}\,\nu_{\bar{\tau}}$, we sum up the energy rate for each
single flavor,
\begin{eqnarray} \label{eq:Me1}
\dot{Q} &=& \dot{Q}_{\nu_e\,\bar{\nu}_e} +
\dot{Q}_{\nu_\mu\,\bar{\nu}_\mu} +
\dot{Q}_{\nu_\tau\,\bar{\nu}_\tau}\nonumber\\
&=& \frac{28 G_F^2 \pi^6 \zeta(5)}{9 c^5 h^6} (D_1 +
2D_2)R_\nu^3(kT_\nu)^9.
\end{eqnarray}

\begin{figure}[htb]
\begin{center}
\epsfig{file=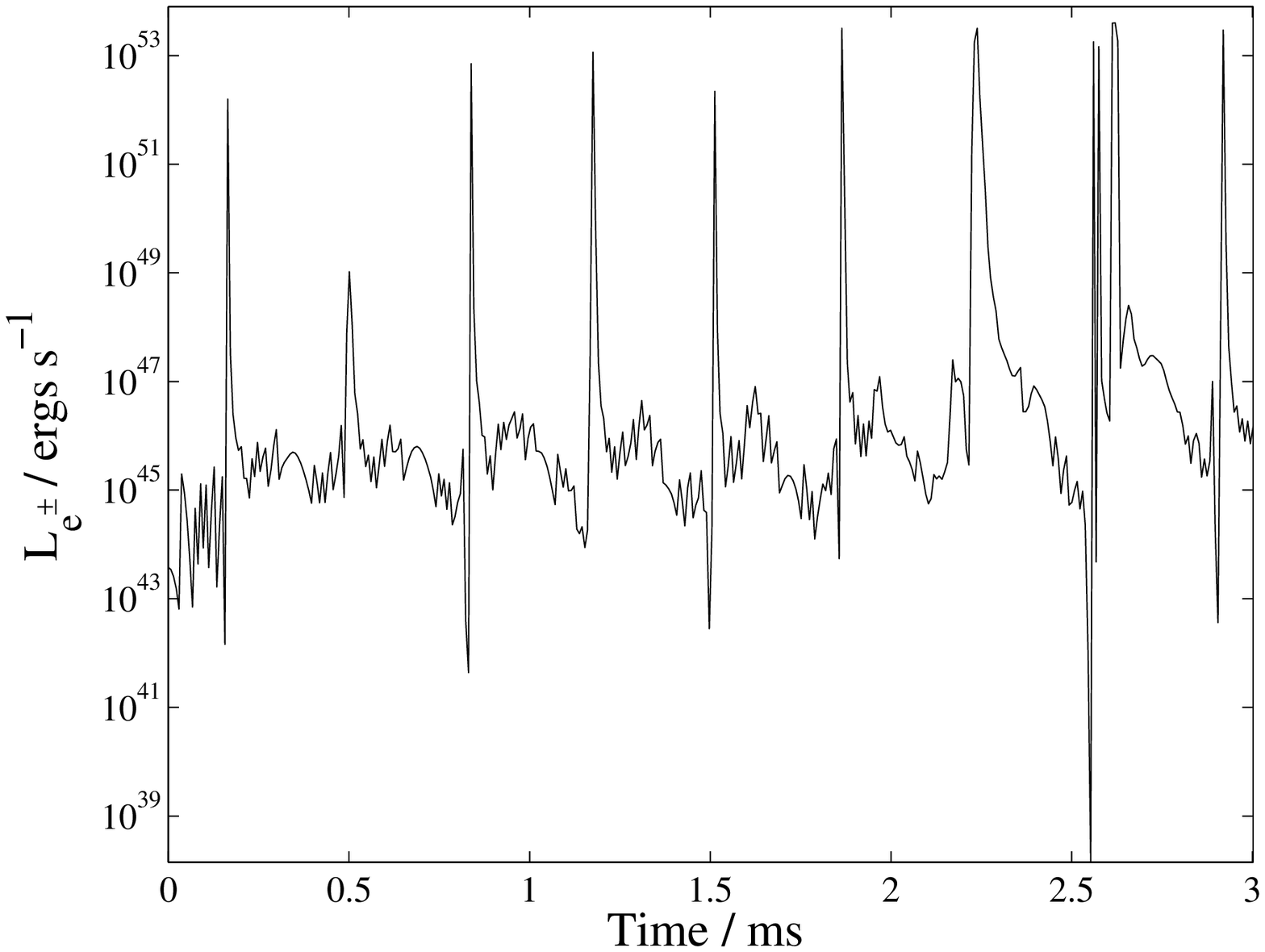,height=2in}
\epsfig{file=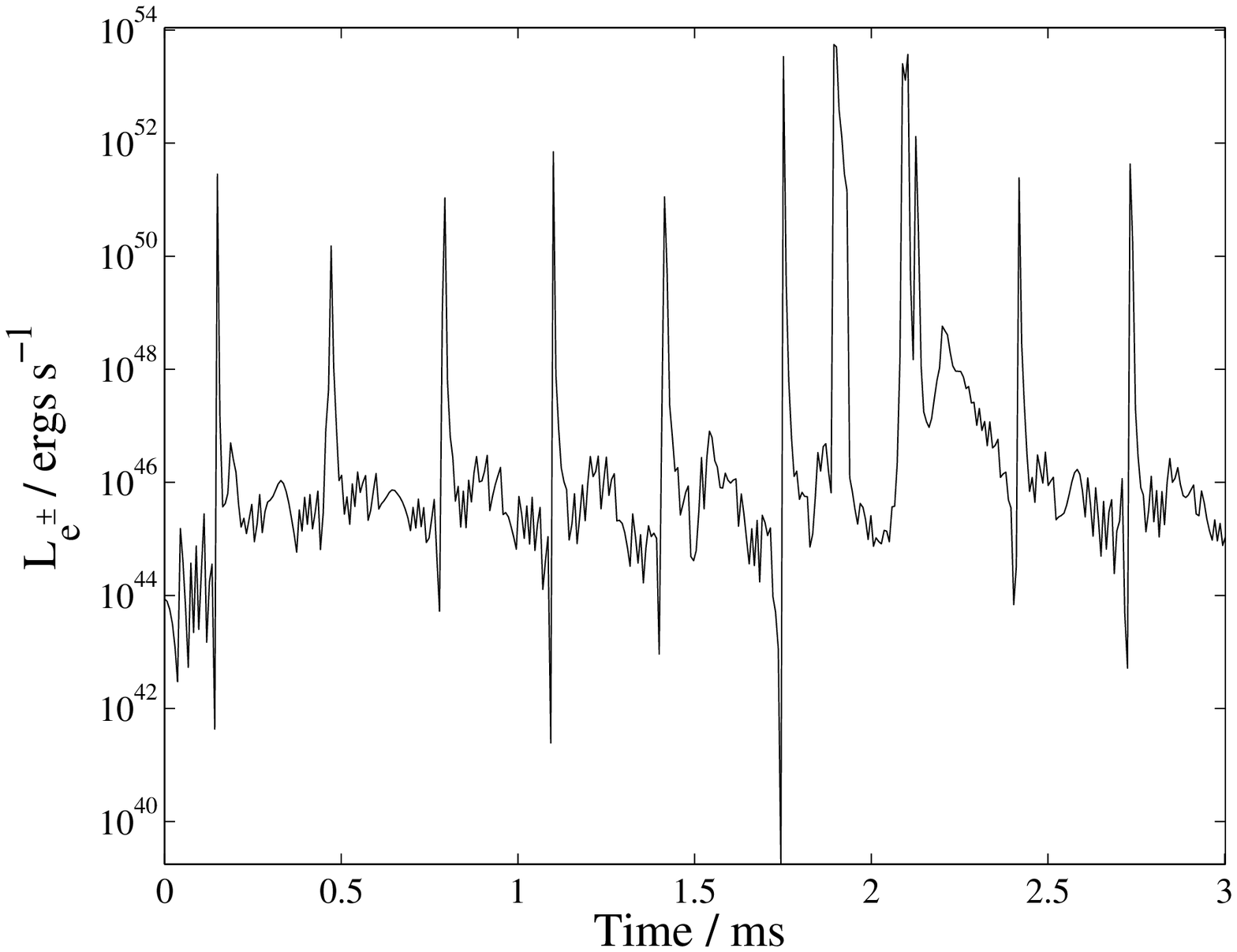,height=2in}
\epsfig{file=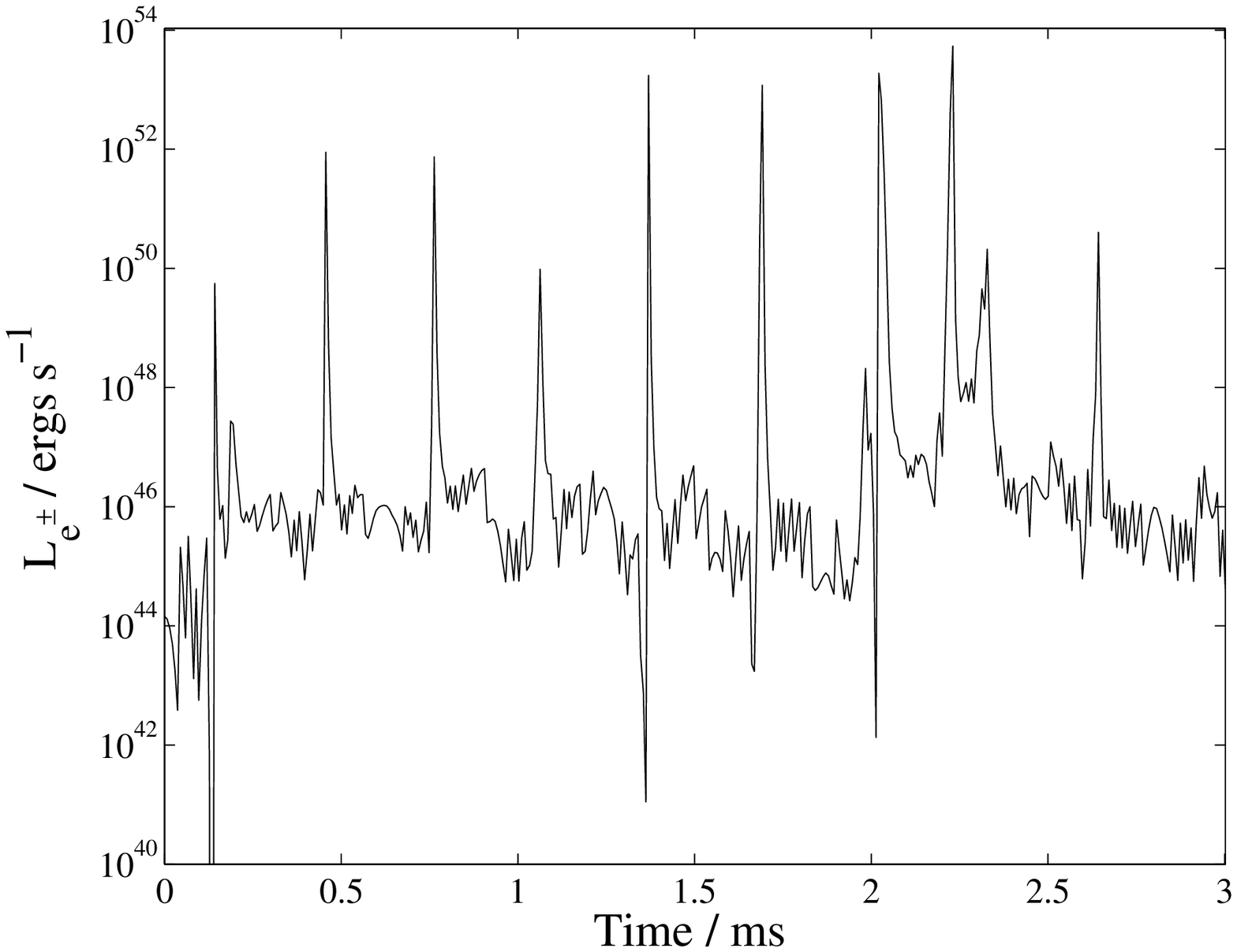,height=2in}
\epsfig{file=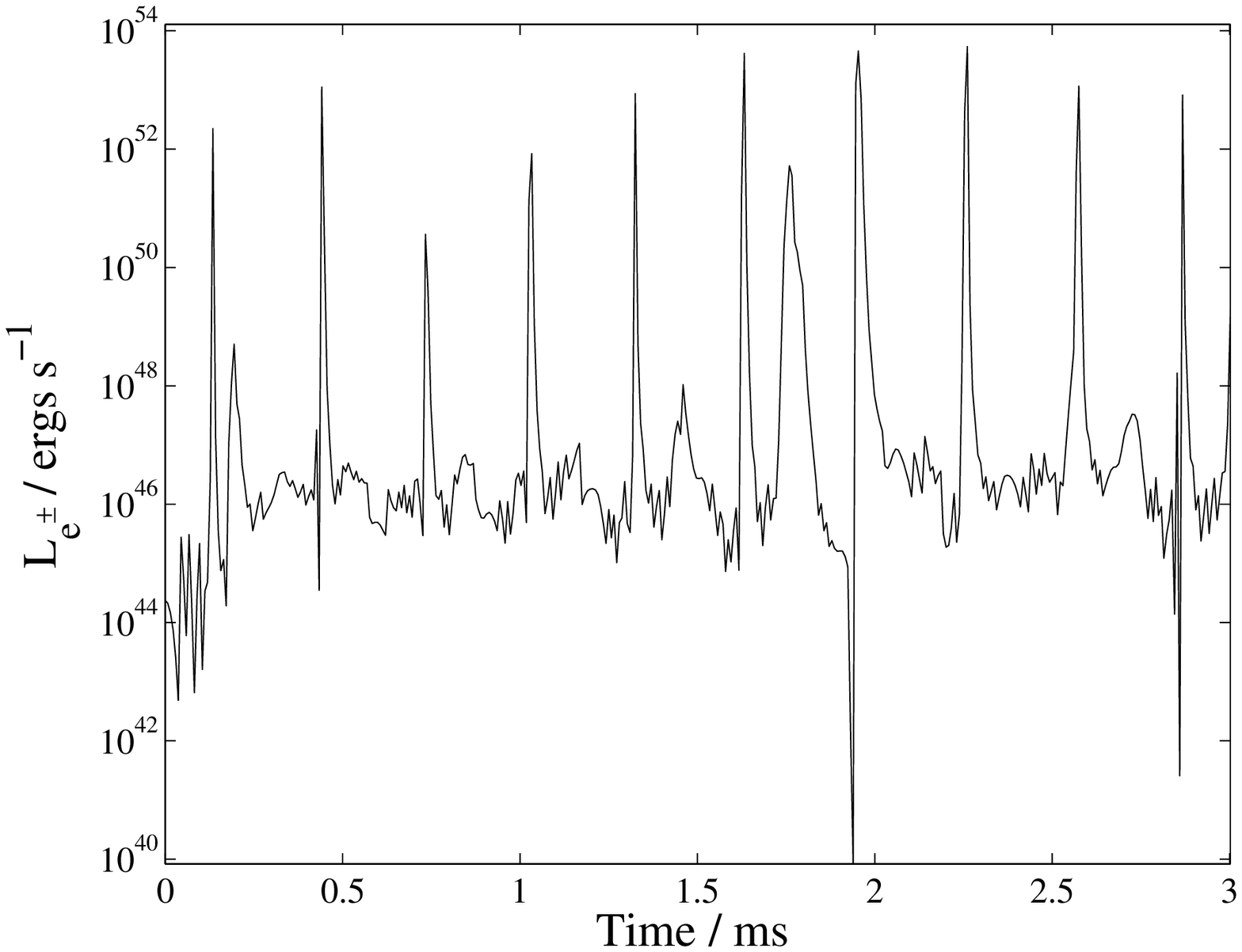,height=2in}
\caption{Electron-positron luminosity versus time for $M=1.55M_{\odot}$, (left upper figure), $M=1.7M_{\odot}$, (right upper figure), $M=1.8M_{\odot}$, (left lower figure), $M=1.9M_{\odot}$, (right lower figure). }
\label{fig:Le_time_models}
\end{center}
\end{figure}

 Fig.~\ref{fig:Le_time_models} shows the rate of energy carried away by the
electron/positron pairs produced through neutrino annihilation,
which varies from $\sim 10^{51}$ergs/s to $\sim 10^{53}$ergs/s.

It is interesting to note that almost all neutrinos can be annihilated
into electron-positron pairs at the peak because of the extremely
high density and high energy of the neutrinos.  In particular the
rest mass of the electrons/positrons is much smaller than
$kT_{\nu}$.


\section{Discussions and final remarks}

In this paper we have studied the possible consequences of the
phase-induced collapse of neutron stars to strange stars. We have
found that both the density and the temperature inside the star will
oscillate with the same period, but almost 180$^{\circ}$ out of
phase, which will result in the emission of intense pulsating
neutrinos and  pairs.
 We want to point out that the intense pulse neutrino luminosity can be maintained due to the oscillatory fluid motion, which
can carry thermal energy directly from the stellar core to the surface. This process can replenish the energy loss of neutrino
emission much quicker than the neutrino diffusion process.
A large fraction of the neutrino energy,
roughly (1-1/$e$), will be absorbed by the matter very near the
stellar surface. When this amount of energy exceeds the
gravitational binding energy, some mass near the stellar surface
will be ejected, and this mass will be further accelerated by
absorbing pairs created from the neutrino and antineutrino
annihilation processes outside the star. Although matter will be
ejected periodically, each ejecta can have different masses and
Lorentz factors, and therefore the intrinsic period could not be
observed. We suggest that the collisions among these ejecta may
produce short GRBs. Our numerical simulations are simulating a
spherically symmetric and non-rotating  collapsing stellar object,
and they also do not contain magnetic fields. Therefore the
radiation emission produced in this model is isotropic. However, a
realistic neutron star should have finite angular momentum and
strong magnetic field, and hence these two factors could produce
asymmetric mass ejection. This effect will be considered in future
work.

The phase-transition from a neutron
star to a strange star was simulated in \cite{FrWo98}, with the conclusion that this process is most
likely not a gamma-ray burst mechanism. They mimic the
phase-transition by the arbitrary motion of a piston deep within the
star, and they have found that the mechanic wave will eject $\sim
10^{-2}M_{\odot}$ baryons, which causes the baryon contamination for
the gamma-ray bursts. In our simulations, we assume a sudden change
of equation of state to mimic the phase-transition, and we use the
Newtonian hydrodynamic code to study the response of the stellar
interior after such a sudden change of the EOS. In our simulations
we find that the mass ejection by the motion of the fluid is very
small. We estimate that the major mass ejection would result from
the heating of neutrinos and pairs on the stellar crust, which is
not modeled in the simulations. Our total energy output and total
mass in ejecta are close to that of \cite{FrWo98}.
However, the neutrino energy injection is pulsating, and hence the
mass ejection is also pulsating. The mass of individual ejecta range
from $\sim 10^{-9}$ to $\sim 10^{-4}M_{\odot}$, with output energy
in the range of $10^{48}$ to $10^{50}$ ergs. Therefore, some ejecta
cannot be relativistic, and they cannot contribute to GRBs.
However,
there are still many relativistic ejecta in each simulation model, which
can have Lorentz factors $>$100, and with a total energy of $\sim
10^{50}$ --- $10^{51}$ ergs.
This could be a possible mechanism for short GRBs.

\bigskip
KSC and TH are supported by the GRF Grants of the
Government of the Hong Kong SAR under HKU7013/06P and HKU7025/07P
respectively.






\end{document}